\documentstyle[prl,aps,twocolumn,epsfig]{revtex}
\newcommand{\eps}{\varepsilon}
\newcommand{\Li}{{\rm Li}}
\newcommand{\Cl}{{\rm Cl}}
\newcommand{\replabel}[1]{\hbox{\raise 12pt\hbox to\hsize{\hfill\rm #1}}}
\begin{document}
\draft
\preprint{MZ-TH/99-13, CLNS/99-1613}
\title{\replabel{MZ-TH/99-13, CLNS/99-1613, hep-ph/9904304, April 1999}
  Transcendental numbers and the topology of three-loop bubbles}
\author{S.~Groote,$^{1,2}$ J.G.~K\"orner$^1$ and A.A.~Pivovarov$^{1,3}$}
\address{$^1$ Institut f\"ur Physik, Johannes-Gutenberg-Universit\"at,
  Staudinger Weg 7, D-55099 Mainz, Germany\\
  $^2$ Floyd R.~Newman Laboratory of Nuclear Studies,
  Cornell University, Ithaca, NY 14853, USA\\
  $^3$ Institute for Nuclear Research of the
  Russian Academy of Sciences, Moscow 117312}
\maketitle
\begin{abstract}
We present a proof that all transcendental numbers that are needed for the
calculation of the master integrals for three-loop vacuum Feynman
diagrams can be obtained by calculating diagrams with an even simpler
topology, the topology of spectacles.
\end{abstract}

Feynman diagrams belong to the basic objects needed in the study of present
phenomenological elementary particle physics. They provide a simple and
convenient language for the bookkeeping of perturbative corrections
involving many-fold integrals. In order to arrive at physical predictions
the many-fold integrals have to be evaluated explicitly. If one is only
aiming at practical applications a numerical evaluation of the integrals
may be sufficient. However, their analytical evaluation is theoretically
more appealing. This line of research has been vigorously pursued during
the last few decades. Major breakthroughs were marked by the introduction
of the integration-by-parts technique~\cite{ibyparts} and the algebraic
approach~\cite{b3b4} which lead to an intensive use of computers for the
necessary symbolic calculations. The new techniques allow one to
systematically classify the structure of integrands and to undertake the
rather massive computations necessary for multi-loop calculations which may
include tens of thousands of Feynman diagrams (see e.g.\
refs.~\cite{g4,b4g4}). Still the analytical evaluation of the basic prototype
integrals remains an art and even has given rise to a new field of research,
namely the study of the transcendental structure of the results as well as
their connection to topology and number theory (see e.g.\
refs.~\cite{dirk1,brograkre,kreimer}).

In general, Feynman integrals are not well defined and require regularization.
The dimensional regularization scheme is the regularization scheme favoured
by most of the practitioners. In contrast to the Pauli-Villars
regularization scheme and other subtraction techniques, the dimensional
regularization method is the most natural one in the sense that it preserves
various features of the diagrams. For example, massless diagrams (i.e.\ with
vanishing bare mass) remain massless under dimensional regularization.

The transcendental structure of massless multi-loop integrals is rather
well understood. It is expressible mainly in terms of Riemann's
$\zeta$-function, $\zeta(L)$ where $L$ depends on the topology of the
diagram and the number of loops. In contrast to this, the massive case is
more complicated and has a richer transcendental structure. At present this 
field is being very actively studied and unites the community of
phenomenological physicists and pure mathematicians. Very recently some
novel and striking results concerning three-loop vacuum bubbles have been
discovered in this field~\cite{broad}.

The integration-by-parts technique within dimensional
regularization~\cite{ibyparts} reduces the calculation of a general
three-loop vacuum diagram to several master configurations. This
reduction involves only algebraic manipulations and is universal for any
given space-time dimension $D$ (see e.g.\ ref.~\cite{b3b4}). A general
strategy for reducing all three-loop vacuum diagrams to a finite set of
fixed master integrals through recurrence relations was described in
ref.~\cite{avdeev}. The analytical structure of the remaining unknown master
integrals with tetrahedron topology has been identified with the help of
ultraprecise numerical methods~\cite{broad}. These new achievements allowed
one to obtain the complete analytical expression for the three-loop
$\rho$-parameter of the Standard Model which was known before
only numerically~\cite{broad,rho,rhop}. The results can be written in terms
of a finite set of transcendental numbers called primitives. Which of these
primitives enter the final result for a particular diagram depends on how
the masses are distributed along the lines of a specific diagram.

The main objects of the calculation in~\cite{broad} were the finite parts
$F_i$ of the tetrahedron diagrams (we adopt the notation of
ref.~\cite{broad}). The diagrams were considered in four-dimensional
space-time while the overall ultraviolet divergence appears as a simple
pole in $\eps=(4-D)/2$ within dimensional regularization. By extensive
use of ultra-high precision numerical calculations it was found that only
two new transcendental numbers $U_{3,1}$, $V_{3,1}$ and the square of
Clausen's dilogarithm $\Cl_2(\theta)={\rm Im}(\Li_2(e^{i\theta}))$ (for
discrete values of its argument) enter the final results. The presence of
the square of Clausen's dilogarithm $\Cl_2(\theta)^2 $ (being the square of
Clausen's dilogarithm appearing already at the two-loop level) had been
conjectured on the basis of the assumption that the primitives form an
algebra~\cite{broad}. The quantity $U_{3,1}$ is related to the master
integral $B_4$ found earlier and is expressible through the polylogarithm
value $\Li_4(1/2)$~\cite{b3b4}. The quantity $V_{3,1}$ which emerges in the 
analysis of vacuum diagrams with tetrahedron topology appears to be
entirely new~\cite{broad}.

In the present note we discuss some new results obtained in the field of
three-loop vacuum bubble diagrams. We show how to identify the above
mentioned primitives in the simpler spectacle topology or in the even
simpler water melon topology (see e.g.\ ref.~\cite{xrec}). Our calculations
are done within two-dimensional space-time where the ultraviolet divergences
are less severe and the integrals and final results are simpler. By keeping
masses finite we have good infrared behaviour. Our previous results tell us
that the transcendental structure for the water melon topology in two
dimensions is the same as in the case of four-dimensional
space-time~\cite{xrec,wm}. There is strong evidence that this is also true
for other topologies and for any even space-time dimensionality.

\begin{figure}[htb]\begin{center}
{\epsfig{figure=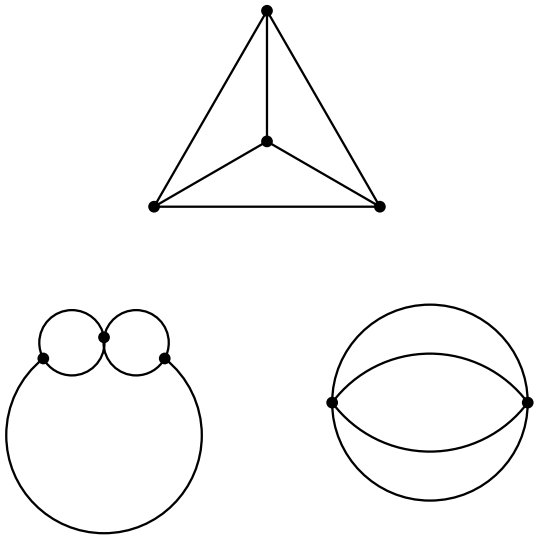, height=7cm, width=7cm}}
\vspace{12pt}\end{center}
\caption{The three three-loop vacuum bubble topologies that are the subject
of this note: The tetrahedron topology (top), the spectacle topology (bottom
left), and the water melon topology (bottom right)}
\end{figure}
Diagrams with tetrahedron topology (Fig.~1) have been analyzed in~\cite{broad} 
using arbitrary combination of massless and massive lines albeit with a single
mass scale $m$. Analytical results for all possible mass configurations
containing only few transcendental numbers have been obtained with the help 
of ultra-high precision (thousands decimal points) numerical
calculations~\cite{broad}. The key observation presented in this note is
that all necessary transcendental numbers that appear in the tetrahedron
case can already be found in the simpler spectacle and water melon
topology shown in Fig.~1. These topologies are sufficiently simple to allow
one to perform all necessary integrations analytically. We present analytical
results for the three-loop spectacle and water melon diagrams, i.e.\ their
transcendentality structure, without ever using any numerical tools. 

The main building block for the treatment of three-loop vacuum bubbles is
the one-loop two-line massive correlator $\Pi(p^2)$ in $D=2-2\eps$
dimensional (Euclidean) space-time,
\begin{eqnarray}\label{exphyper}
\lefteqn{\Pi(p^2)\ =\ \int\frac{d^Dk}{((p-k)^2+m^2)(k^2+m^2)}}\\
  &=&\frac{2^{3+2\eps}\pi^{1-\eps}\Gamma(1+\eps)}{(p^2+4m^2)^{1+\eps}}\
  {}_2F_1\left(1+\eps,\frac12;\frac32;\frac{p^2}{p^2+4m^2}\right)\nonumber
\end{eqnarray}
with $ {}_2F_1(a,b;c;z)$ being the hypergeometric function. An alternative
representation of the correlator is obtained through the dispersion relation
\begin{equation}\label{disp}
\Pi(p^2)=\int_{4m^2}^\infty\frac{\rho(s)ds}{s+p^2}
\end{equation}
with
\begin{equation}\label{specden}
\rho(s)=\frac{(s-4m^2)^\eps}{2\pi\sqrt{s(s-4m^2)}}\
  \frac{\pi^{1/2-\eps}}{2^{4\eps}\Gamma(1/2+\eps)}.
\end{equation}
In order to reproduce the transcendental structure of the finite parts of the
tetrahedron in four dimensions we need a first order $\eps$ expansion of
water melons and spectacles near two-dimensional space-time. Note that
these diagrams are well-defined and ultraviolet finite in two dimensions
and, formally, require no regularization. However, the sought-for
transcendental structure appears only in higher orders of the $\eps$
expansion while the leading order is simple and contains only the standard
transcendental numbers such as $\zeta(3)$ or $\ln(2)$. Therefore we write 
\begin{equation}
\Pi(p^2)=\Pi_0(p^2)+\eps\Pi_1(p^2)+O(\eps^2)
\end{equation}
and keep only the first order in $\eps$ which happens to be sufficient for
our goal of finding all the transcendental numbers appearing in the
tetrahedron case. Using either the explicit formula in Eq.~(\ref{exphyper})
or the dispersive representation Eq.~(\ref{disp}) with the spectral density 
given by Eq.~(\ref{specden}) and expanded to the first order in $\eps$, we
find
\begin{equation}
\Pi_0(4m^2\sinh^2(\eta/2))={\eta\over 4\pi m^2 \sinh{\eta}}
\end{equation}
where the variable $\eta$ has been introduced for convenience,
$\sqrt{p^2}=2m\sinh({\eta/2})$. In the first order of the $\eps$ expansion
we have
\begin{equation}
\Pi_1(4m^2\sinh^2(\eta/2))={f(e^{-\eta})\over 4\pi m^2 \sinh{\eta}}  
\end{equation}
with 
\begin{eqnarray}\label{ft}
f(t)&=&2\Li_2(-t)+2\ln t\ln(1+t)-\frac12\ln^2(t)+\zeta(2)\nonumber\\
  &=&2\int_0^t\frac{\ln u}{1+u}du-\frac12\ln^2t+\zeta(2).
\end{eqnarray}

The integral for the water melon diagram is given by
\begin{equation}\label{wm11}
W=2\pi m^2\int\Pi(p^2)^2d^2p=W_0+\eps W_1+O(\eps^2).
\end{equation}
Note that here and later we use a two-dimensional integration measure. This
prescription differs from the standard dimensional regularization but
suffices for our purposes and makes the final expressions simpler. The use
of a $D$-dimensional integration measure would not change the functional
structure of the integrands and would simply generate some additional terms
that can be analyzed within the same technique. Upon expanding the integral
in Eq.~(\ref{wm11}) in powers of $\eps$ we obtain the leading term 
\begin{equation}\label{WM00}
W_0=2\pi m^2\int\Pi_0(p^2)^2d^2p=\frac78\zeta(3)
\end{equation}
and the first order term
\begin{equation}\label{finalW}
W_1=4\pi m^2\int\Pi_0(p^2)\Pi_1(p^2)d^2p=\int_0^1\frac{2\ln t}{1-t^2}f(t)dt.
\end{equation}

The spectacle diagram is given by the integral 
\begin{equation}\label{finalS}
S=2\pi m^4\int\frac{\Pi(p^2)^2}{p^2+M^2}d^2p=S_0+\eps S_1+O(\eps^2),
\end{equation}
where the single ``frame'' propagator has a mass $M$ which differs from the
other mass parameter $m$ in the ``rim'' propagators. The expression for the
leading order term $S_0$ is simple (see ref.~\cite{xrec}) while the first
order term $S_1$ (which is of interest for us here) reads
\begin{eqnarray}\label{spe11}
S_1&=&\int_0^1\frac{2t\,f(t)\ln t\,dt}{(1-t^2)(t^2-2t\cos\theta+1)}\nonumber\\
  &=&\int_0^1\frac{2t\,f(t)\ln t\,dt}{(1-t^2)(\lambda_0-t)(\bar\lambda_0-t)}
\end{eqnarray}
where $\lambda_0=e^{i\theta}$, $\cos\theta=1-M^2/2m^2$.

By partial fractioning the rational expressions in the integrands of
Eqs.~(\ref{finalW}) and~(\ref{spe11}) we find that the most complicated
integral in both cases has the form
\begin{equation}
\int_0^1\frac{\ln t}{\bar\lambda-t}f(t)dt
  = 2I(\lambda)+3\Li_4(\lambda)-\zeta(2)\Li_2(\lambda)
\label{basicI}
\end{equation}
where $I(\lambda)$ is a generic nonreducible term which cannot be expressed 
with the help of the transcendentality structure that occured earlier on.
One has 
\begin{equation}
I(\lambda)=\int_0^1dt\frac{\ln t}{\bar\lambda-t}\int_0^tdu\frac{\ln u}{1+u}
\label{gene}
\end{equation}
while the integral of the last two terms in Eq.~(\ref{ft}) is explicitly
expressed through polylogarithms $\Li_2$ and $\Li_4$. For the relevant
values of $\lambda$ this generic integral in Eq.~(\ref{gene}) contains all
the primitives $U_{3,1}$, $V_{3,1}$ and also Clausen's polylogarithms
$\Cl_2$ and $\Cl_4$. 

The value $\lambda=1$ occurs in both water melon and spectacle cases. For
this value we obtain
\begin{equation}
I(1)=\frac{17\pi^4}{1440}+2U_{3,1}
\end{equation}
with the explicit expression for the primitive $U_{3,1}$~\cite{broad}
\[
U_{3,1}=\frac12\zeta(4)+\frac12\zeta(2)\ln^2(2) 
-\frac{1}{12}\ln^4(2)- 2 \Li_4\left(\frac12\right)
\]
where $\zeta(4)=\pi^4/90$ and $\zeta(2)=\pi^2/6$.

The part present in the spectacle diagram in Eqs.~(\ref{finalS})
and~(\ref{spe11}) depends on the mass ratio. For $M=m$ we have
$\theta=\pi/3$, so $\lambda_0=e^{i\pi/3}$ is one of the sixth order roots
of unity. This observation discloses the special role of the sixth order
roots of unity which had been observed before in ref.~\cite{broad}. For
$\lambda=e^{i\pi/3}$ we obtain
\begin{eqnarray}
I(e^{i\pi/3})&=&\frac{197\pi^4}{38880}-\frac13\Cl_2^2\left(\frac\pi3\right)
  +2V_{3,1}+\frac{5i\pi^3}{162}\ln 3\nonumber\\&&
  +\frac{13}{108} i\pi^2\Cl_2\left(\frac{\pi}{3}\right)
  -\frac{35i}{18}\Cl_4\left(\frac\pi3\right).
\end{eqnarray}
The primitive $V_{3,1}$ is given by 
\begin{equation}\label{primV}
V_{3,1}=\sum_{m>n>0}(-1)^m\cos\left(\frac{2\pi n}3\right)\frac1{m^3 n}.
\end{equation}
In the case $M=\sqrt3m$ we end up with $\lambda_0=e^{2i\pi/3}$, another
sixth order root of unity. For this value of $\lambda$ we obtain
\begin{eqnarray}
I(e^{2i\pi/3})&=&-\frac{79\pi^4}{12960}
  +\frac13\Cl_2^2\left(\frac\pi3\right)\nonumber\\&&
  +\frac{7i\pi^2}{36}\Cl_2\left(\frac\pi3\right)
  -\frac{11i}{6}\Cl_4\left(\frac\pi3\right).
\end{eqnarray}
This expression is simpler and does not contain the new primitive $V_{3,1}$.

The case $M=2m$ is a degenerate one, $\lambda_0=e^{i\pi}=-1$ and the
expression for $I$ reduces to $\zeta$-functions only,
\begin{eqnarray}\label{degene}
I(-1)&=&-\frac{\pi^4}{288}=-\frac5{16}\zeta(4).
\end{eqnarray}

Finally, the complete expression for the spectacle (or its most interesting
part $S_1$) can be easily found by collecting the above results. For the
standard arrangement of masses $M=m$ we have  
\begin{eqnarray}
S_1&=&-\frac{251\pi^4}{58320}+4U_{3,1}-\frac{16}{3} V_{3,1}
+  \frac{8}{9}\Cl_2^2\left(\frac\pi3\right).
\label{spfi1}
\end{eqnarray}

We have thus fulfilled our promise and have discovered all magic numbers
$U_{3,1}$, $V_{3,1}$ and $\Cl_2^2(\pi/3)$ which one encounters in the
evaluation of tetrahedron vacuum diagrams using different combinations of
massless and massive lines with a single mass scale $m$. We have found the
magic numbers in the simpler three-loop water melon and spectacle
topologies. The reason for this accomplishment can be read off from our
analytical expressions: all integrals (or the functional structures of
integrands) which appeared in the calculation of the finite parts of the
tetrahedron diagrams are contained in our basic object $I(\lambda)$. We see
the origin of the important role played by the sixth order root of unity:
for a single mass $m$ this number appears as a root of the denominator in
the spectacle diagram. Within the analysis presented in ref.~\cite{broad}
the special role of this sixth order root of unity had no rational
explanation. Having discovered this, we extended the analysis to arbitrary
values of $\lambda$ in Eqs.~(\ref{spe11}), (\ref{basicI}) and~(\ref{gene})
by introducing a second mass parameter $M$ in the spectacle diagram given
by Eq.~(\ref{finalS}). Concerning possible future extensions of our approach
we emphasize that the appearance of the relevant transcendental
structure at the level of the simpler topologies is quite an essential
simplifying feature if one wants to proceed to even higher-loop
calculations. For example, the simplicity of the water melon topology makes
them computable with any number of loops~\cite{wm}. In this sense the
present calculations can be considered to be a first step towards the
evaluation of four-loop vacuum bubbles. Our result finally leads us to a
conjecture about the calculability of the three-loop master integrals. If
they are reducible to spectacle and water melon diagrams at the analytical
level, this observation may lead to a way beyond the time-consuming
integration-by-parts technique for the evaluation of three-loop diagrams.

To conclude, by analyzing massive three-loop vacuum bubbles belonging to
the spectacle topology class within dimensional regularization for
two-dimensional space-time, we discovered and identified analytically all
transcendental numbers which were previously found by numerical methods for
the three-loop tetrahedron topology.  

\acknowledgments
The work is supported in part by the Volkswagen Foundation under contract
No.\ I/73611. A.A.~Pivovarov is supported in part by the Russian Fund for
Basic Research under contracts Nos.~97-02-17065 and 99-01-00091. S.~Groote
gratefully acknowledges a grant given by the Max Kade Foundation.

\end{document}